\documentclass[10pt,a4paper]{article}
\usepackage[english]{babel}
\begin{document}
\begin{center}
{\bf \Large
Mechanical and Dielectric Response of PZT and 
Tetragonal to Monoclinic Phase Transition: I. Theory 
}
\end{center}
\vskip3cm 
\begin{center}
O. Hudak

Istituto dei Sistemi Complessi CNR, Via del Fosso del Cavaliere, Roma 
\end{center}
\begin{center}

F. Cordero

Istituto dei Sistemi Complessi CNR, Via del Fosso del Cavaliere, Roma 

\end{center}
\begin{center}

F. Craciun

Istituto dei Sistemi Complessi CNR, Via del Fosso del Cavaliere, Roma 

\end{center}

\newpage
\section*{Abstract.}
The PZT material near the morphotropic phase boundary where a monoclinic 
phase was observed is studied in this paper. A theory of the tetragonal to monoclinic phase transition was developed within the Landau free energy 
approach. The order parameter is the electric polarization vector. The phase 
transition from the tetragonal to the monoclinic phase is of the first order near the second 
order. The tetragonal and monoclinic phases are described as equilibrium states 
within this theory. Dielectric response near the phase transition from the tetragonal to the monoclinic phase was studied for the crystal and for the ceramic and polycrystalline 
material taking into account the diagonal 33 and the off diagonal 31 
components of susceptibility. The later diverges at 
temperature of stability boundary for the monoclinic phase, and it is zero in 
the tetragonal phase. Such a behavior is not observed in the effective dielectric 
static constant temperature dependence for ceramics and polycrystalline materials. 
We explain absence of a large contribution of the off diagonal susceptibility 
by a grain structure of the material and by percolation properties of this microcomposite. 
Above the percolation transition the off diagonal contribution to 
the effective susceptibility is not present. Observed temperature behavior of the 
effective dielectric susceptibility is thus qualitatively explained. Effective 
susceptibility which is decreasing in tetragonal phase is in the monoclinic phase also decreasing, 
however below phase transition temperature the decrease is slower. A quantitative theoretical 
behavior of dielectric susceptibility for both phases is found. Mechanical 
response of PZT material to a uniaxial stress is studied theoreticaly for both phases. Observed 
temperature behavior of it, a well around the transition temperature from the tetragonal to monoclinic phase transition, is qualitatively 
explained.

\newpage
\section{Introduction.}

PZT materials are interesting ferroelectric materials studied for many years, \cite{1} to \cite{4}. Near $x \approx  0.48$ a morphotropic phase boundary 
(MPB) exists \cite{5}. It is a region between the tetragonal phase, 
which is on the Ti rich side, and the rhombohedral phase, which is on the Zr rich 
side. This region is associated with many interesting dielectric, piezoelectric and 
other properties of PZT \cite{6}. The width of MPB varies with the homogeneity 
of the powders \cite{7} to \cite{9} and the size of grains \cite{10}. Recently a new phase 
was found around MPB in the phase diagram. This phase is a monoclinic phase \cite{5}. Measurements of the elastic response \cite{11} on PZT have shown the presence of 
the transition from the cubic to tetragonal phase and a well at lower temperatures. 
The transitions from the cubic to the tetragonal and from the cubic 
to the rhombohedral phases are understood quite well, see in \cite{5}. A transition 
from the tetragonal to the monoclinic phase which is not well understood will be 
studied in this paper. To describe the tetragonal to the monoclinic phase 
transition we need to identify the order parameter and then to use the Landau 
theory of phase transitions, this transition is of the first order near the second 
order. Dielectric response and mechanical response of PZT will be studied for 
the monoclinic phase and for this phase transition in this paper. We compare 
our results with experimental results, \cite{11} and \cite{12}. These materials are ceramics 
and polycrystals, they are composites and this fact does lead to a modification 
of the dielectric response as it is in the case of some ferroelastic­dielectric microcomposites 
\cite{26}. Crystallites with different phases, tetragonal and monoclinic or rhombohedral, will not be considered here. The effect of different phases coexisting in 
PZT was considered in \cite{13}.

The paper has the following sections. In the following section we describe 
briefly PZT and the Morphotropic Phase Boundary problem. Then we study 
a phase transition from the tetragonal phase to the monoclinic phase. Transition temperature region 
between the monoclinic and the tetragonal phase is studied in more details. 
This enables us to study dielectric response near the phase transition. PZT material 
is usually ceramics or polycrystalline. Then its microcomposite properties 
influence the dielectric response which we study in the next section. Mechanical 
response of PZT near the phase transition from the tetragonal phase to the monoclinic phase is 
studied theoreticaly then. Dielectric and mechanical responses theoreticaly described 
in this paper for this temperature range qualitatively describe observed 
dependencies of them on temperature. In the last section there is a summary 
of obtained results in this paper.

\section{PZT and the Morphotropic Phase Boundary.}

The morphotropic phase boundary (MPB) in PZT is not well understood \cite{5}. 
There is evidence that this phase boundary is a two phase region, see in \cite{13} and \cite{14}. There are also measurements which show that this is a single 
phase region, or part of it is a two phase region (tetragonal and  monoclinic)and 
part single phase region, see in \cite{13} where the author addresses the question of 
coexistence of different crystallographic phases within the same crystallite. 
Materials similar to PZT which have MPB are PMN PT and PZN PT \cite{13}. It 
was found in \cite{14} that when the diffusion of Zr and Ti atoms in PZT is fast 
enough a two phase region exists at this boundary. When the diffusion is 
not fast enough a nonequilibrium region forms and the classical phase diagram 
\cite{12} appears in this material.

To understand the nature of the MPB we will use the Landau theory for phase transitions. As it is noted in \cite{13} the monoclinic phase 
as a intermediate phase between the tetragonal and the rhombohedral phase may 
exists and that the tetragonal space group P4mm and the rhombohedral space group 
R3m both have as a subgroup the monoclinic space group Cm . Thus there exists \cite{13} the possibility that a tetragonal to rhombohedral phase transition could take 
place via a monoclinic intermediate state. It is then very usefull to understand 
properties of the monoclinic phase and of the phase transition from tetragonal 
to monoclinic phase. In \cite{15} the authors have concluded based on local electron 
diffractions, that the local structure of PZT is always monoclinic, and that the 
average tetragonal and rhombohedral structures are a result of short­range to 
long­range ordered states. According to these authors these viewpoints may 
stimulate new constitutive models of ferroelectrics with compositions close to a 
MPB, and that so far, a constitutional model which deals with changes of the 
structural order is not available \cite{15}. 
It is then very useful to understand properties of the monoclinic phase and 
of the phase transition from the tetragonal phase to the monoclinic phase. First principles 
calculations \cite{16} have shown the stability of the monoclinic phase in a narrow 
region of MPB when a random Zr/Ti cation distribution was taken into account. 
The stability of a monoclinic phase of space group Cm in between the rhombohedric 
and tetragonal phases was also explained using a Landau­Devonshire 
approach \cite{17}. The authors of \cite{17} predicted existence of two other monoclinic phases between 
the rhombohedral and orthorhombic phases, and the orthorhombic and tetragonal 
phase of perovskites, respectively. The later authors have found that the 
Landau theory of the sixth order in order parameter for the transition from the cubic to the tetragonal and to the rhobohedral phase cannot explain presence of 
a monoclinic phase and thus they extend the theory to the eight order and then 
they are able to explain presence of a monoclinic phase. The monoclinic phase 
of PZT ceramics, Raman and phenomenological theory studies were done 
in \cite{18}. The authors \cite{18} have found using Raman scattering the tetragonal to 
monoclinic phase transition in PZT ceramics near morphotropic phase boundary 
at low temperatures. They have found that the transition is characterized by 
changes in the frequency of lattice modes with the temperature. To discuss the 
stability of the monoclinic phase the authors \cite{18} used the Landau Devonshire 
phenomenological theory. While these authors and \cite{19} to \cite{23} were able to predict 
the rhombohedric to tetragonal phase transition at approximately x=0.5, 
their calculations show that the monoclinic phase is not stable for any value of 
Ti concentration. Free energy as a function of Ti concentration for the rhombohedric, 
tetragonal and monoclinic phases was calculated by these authors at 
­80 degrees of Celsia. According to \cite{18} this result indicates that the free energy 
used in \cite{19} to \cite{23} is not suitable to describe the existence of monoclinic phase. 
To describe the monoclinic phase it is taken in \cite{18} into account an order 
parameter, which is the monoclinic angle. It is noted by these authors that the 
monoclinic angle is not the true order parameter since it does not go to zero at the 
temperature of tetragonal to monoclinic phase transition, and they are using a 
modified order parameter in their Landau theory. They have observed that the 
value of the free energy of the monoclinic phase is now lower than that of free energy 
of the rhombohedric and tetragonal phases in a small region close to MPB 
and they noted that this result shows that the modified free energy correctly 
predicts the existence of a monoclinic phase in the MPB region, in accordance 
with the experimental observations. The vector of the electric polarization is 
according to these authors not a correct order parameter to describe the phase 
transition from tetragonal to the monoclinic phase and they are using the third 
order parameter for the second order phase transition which enable them to describe 
this phase transition. In fact this third order parameter does not follow 
from the symmetry analysis on which the Landau theory of the second order 
phase transition and of the first order near the second order is based. A priori 
order parameter thus does not enable to understand the monoclinic phase and 
the transition from tetragonal to monoclinic phase correctly. We will study the 
monoclinic phase and the transition from tetragonal to monoclinic phase within 
the Landau theory. We note that the free energy expansion found in \cite{19} to \cite{23} 
for the cubic to tetragonal phase transition should be modified by terms dependent 
on the electric polarization and which are invariant in the tetragonal 
phase. The tetragonal to monoclinic phase then will be described correctly. It 
is driven by temperature and the primary order parameter is the electric polarization 
vector. The origin of terms dependent on the electric polarization is in interaction of 
polarization vector with fluctuations of other degrees of freedom which when integrated out in 
the tetragonal phase lead to invariants for this phase. This is consistent with 
the symmetry group analysis and group subgroup relations for possible second 
order and first order near to second order phase transitions, see in \cite{13} and 
\cite{24}. In \cite{13} the author has studied the equations of compatibility, which have 
to be satisfied. He analyzed the three possible phase transitions: tetragonal­ 
rhombohedral, tetragonal­monoclinic, and monoclinic­rhombohedral. He used 
the micromechanics approach. The author found, that the tetragonal rhombohedral 
transition may be compatible at a composition at the very Zr­rich side 
of the MPB $x \approx  0.54$. He notes that in this compositional region there are 
no experimental data for the tetragonal phase, so that it may not exist at this 
Zr concentrations. For the tetragonal to monoclinic phase transition he has found 
that it is never compatible according to the accessible data and that the monoclinic to rhombohedral phase 
transition may be compatible at even higher Zr contents 
as tetragonal rhombohedral. He then discusses the incompatibility between 
the tetragonal and the monoclinic phase, and he has found that it is almost 
constant across the MPB, but relatively low compared to the other transitions. 
This may explain \cite{13}, why tetragonal and monoclinic phases seem to coexist 
within the MPB region.

We will study in our paper the tetragonal and monoclinic 
phases, the transition between them we have found is of the first order 
near the second order. We will not study coexistence of these phases in our 
paper.

The Landau expansion of the Gibbs free energy for the cubic phase was described in \cite{19} to \cite{23} and 
has the form: 
\begin{equation}
\label{1}
F = \int dV [\alpha_{1} (P_{1}^{2} + P_{2}^{2} + P_{3}^{2}) +
\alpha_{11} (P_{1}^{4} + P_{2}^{4} + P_{3}^{4}) +
\alpha_{12} (P_{1}^{2} P_{2}^{2}+ P_{2}^{2} P_{3}^{2} + P_{3}^{2} P_{1}^{2}) +
\end{equation}
\[ + \alpha_{111} (P_{1}^{6} + P_{2}^{6} + P_{3}^{6}) +
\alpha_{112} ( P_{3}^{2}(P_{1}^{4} + P_{2}^{4}) + P_{1}^{2}(P_{2}^{4} + P_{3}^{4}) + P_{1}^{2}(P_{2}^{4} + P_{3}^{4})) + \]
\[+ \alpha_{123} P_{1}^{2}P_{2}^{2}P_{3}^{2}] \]

here $P_{i}$ (i=1,2,3) are the polarization components along the cubic axes, $\alpha_{1}$ is the usual parameter from the Landau theory, $\alpha_{ij} $ (i,j = 1,2) and  $\alpha_{ijk} $ 
(i,j,k=1,2,3) are the dielectric stiffness and high­order dielectric stiffness at constant 
stress, respectively. The coeffcient $\alpha_{1} = 
\alpha_{1,0}(T - T_{0}) $ is a temperature dependent coefficient. Here $T_{0}$ is the transition temperature. The values of all coeffecients in (\ref{1}) and the 
solutions for the tetragonal and rhombohedric phases can be found in \cite{19} to \cite{23}. In the cited papers there is a Landau expansion (\ref{1}) and an expansion in 
other secondary order parameters. Contribution from the elastic tensor and its 
coupling to polarization is neglected in the expansion (\ref{1}). 
While the free energy expansion (\ref{1}) is for the cubic phase, in the case of 
the tetragonal phase \cite{25} the cations displacements lie close to the monoclinic 
$[ \bar{2}01]$ direction, which corresponds to the rhombohedral $[111]$ axis. Thus the 
transition from the tetragonal to the monoclinic phase will be described by the 
free energy from (\ref{1}). This phase transition is of the first order near the second 
order. However note that the free energy (\ref{1}) will contain contributions which 
correspond to the tetragonal phase symmetry now. One of such fourth order terms has the form: 
\begin{equation}
\label{2}
\delta F = \int dV \nu (P_{1}^{2} - P_{2}^{2} )^{2} . 
\end{equation}

The term (\ref{2}) breaks the cubic symmetry, thus it is not present in the corresponding 
free energy expansion (\ref{1}) for the cubic phase. The constant $ \delta $ is not known as concerning its value, we assume that it is not dependent on 
temperature and pressure. However due to the fact that a transition from the 
tetragonal to the monoclinic phase is present in experiments we may assume that it has a very 
large value such that we obtain from the contribution (2) the equality 
$ P_{1}^{2} = P_{2}^{2} $ at the ground state. This condition is used a priori for description of the monoclinic 
phase also in \cite{25}. However its symmetry origin in \cite{25} is not explained. 
The authors of \cite{25} are taking into account a third order parameter in order to 
describe the monoclinic phase, which is the monoclinic angle. This approach is artificial. 
The case of $\delta$ general in (\ref{2}) will be studied elsewhere as well as mechanical 
and dielectric response properties of the system for $\delta$ general. Then 
we may find equations for the polarization components $ P_{1} $ and 
$ P_{3} $ only. Thus 
the order parameter space is two dimensional.

Another of such 
terms has the form: 
\begin{equation}
\label{3}
\delta F = \int dV \tau  P_{3}^{4}.
\end{equation}

The term (\ref{3}) breaks again the cubic symmetry, thus it is not present in the corresponding 
free energy expansion (\ref{1}) for the cubic phase. The constant $\tau $ is not known as concerning its value, we assume that it is not dependent on temperature 
and pressure.  Thus this term prefers the polarization to be oriented 
in the tetragonal axis. However we will not consider this term further in this paper. One may perform a systematic analysis of the free energy expansion corresponding to the symmetry of the tetragonal phase. We decided to use the terms (\ref{1}) and (\ref{2}) in this paper because the coefficients of the terms in (\ref{1}) are well known and the term (\ref{2}) gives zero contribution for the tetragonal phase equilibrium state $P_{3}^{2} > 0$, $P_{1}^{2} = P_{2}^{2} = 0$ and for the monoclinic phase equilibrium state $P_{3}^{2} > 0$, $P_{1}^{2} = P_{2}^{2} > 0$, and it has tetragonal symmetry.
The case of $\delta$ and $\tau$ general will be studied elsewhere as well as corresponding mechanical 
and dielectric response properties.

Decreasing temperature we may 
expect that the Landau theory based on (\ref{1}) and (\ref{2}) describes the transition from the tetragonal to monoclinic phase. In the later phase we will have $P_{1}^{2} = P_{2}^{2} \ddagger P_{3}^{2}$.

\section{Phase Transition: Tetragonal ­ Monoclinic Phase.}

The equations from which we will find the values of the order parameter in the 
equilibrium phase for a given temperature and pressure, i.e. of the polarization components $P_{1}^{2}=P_{2}^{2}$ and $P_{3}$ are: 
\begin{equation}
\label{4}
4 \alpha_{1} P_{1} + 4 ( 2 \alpha_{11} + \alpha_{12} )P_{1}^{3} +
\end{equation}
\[ + 4 \alpha_{12} P_{1}P_{3}^{2} + 12( \alpha_{111}+\alpha_{112})P_{1}^{5} + \]
\[ + 4 \alpha_{112}P_{1}P_{3}^{4} + 
4 ( 2 \alpha_{112} + \alpha_{123} )P_{1}^{3}P_{3}^{2}  = 0. \]

and: 
\begin{equation}
\label{5}
2 \alpha_{1} P_{3} + 4 \alpha_{11} P_{3}^{3} +
\end{equation}
\[ + 4 \alpha_{12} P_{3} P_{1}^{2} + 6  \alpha_{111}P_{3}^{5} + \]
\[+ 8 \alpha_{112}P_{1}^{2}P_{3}^{3} + 2 ( 2 \alpha_{112} + \alpha_{123} )P_{1}^{4}P_{3} = 0. \]

The solution of the equations (\ref{4}) and (\ref{5}) for the cubic phase $P_{1} = P_{2} = P_{3} = 0 $, for the tetragonal phase $P_{1} = P_{2} = 0$ and $ P_{3}^{2} \ddagger  0 $, for the rhombohedric 
phase $P_{1} = P_{2} = P_{3} \ddagger  0 $, and orthorhombic phase $P_{1} = P_{2} \ddagger 0 $ and $ P_{3} = 0 $ can be 
found. However these phases were already studied in \cite{19} to \cite{23}.

In our model due to large constant $\delta$ the polarization components fulfill $P_{1}^{2} = P_{2}^{2} \ddagger  0 $
for the equilibrium state of the monoclinic phase. Let us describe this monoclinic phase. 
To do this we introduce polar coordinates in the order parameter space $(P_{1} , P_{3})$: 
\begin{equation}
\label{6}
P_{1} = A \sin(\phi) 
\end{equation}
\[ P_{3} = A \cos(\phi) \]

where A is the amplitude of the order parameter in this parameter space, and $\phi$ is the angle of the order parameter in this parameter space. Note that 
for the order parameter angle equal to zero, $\phi = 0$, the polarization is oriented 
in the tetragonal direction. For the order parameter angle non­equal to zero, however smaller than $\frac{\pi}{2}$, i.e. for $\frac{\pi}{2} > \phi > 0$, the polarization is oriented in 
the monoclinic direction. For the angle $\phi$ equal to $\frac{\pi}{4}$ the polarization vector is 
oriented in the rhombohedric direction. For the order parameter angle equal to $\frac{\pi}{2}$, the polarization vector is oriented in the orthorhombic direction.

We have found from the equations (\ref{4}) to (\ref{6}) that the amplitude A depends on the phase angle $\phi$ in the following way: 
\begin{equation}
\label{7}
A^{2} = - \frac{1}{r + q \sin^{2}(\phi)} .
\end{equation}

Here r and q constants are defined as: 
\begin{equation}
\label{8}
r = \frac{3 \alpha_{111} - \alpha_{112}}{2 \alpha_{11} - \alpha_{12}},
\end{equation}

and: 
\begin{equation}
\label{9}
q = \frac{2 \alpha_{112} - \alpha_{123}}{2 \alpha_{11} - \alpha_{12}}.
\end{equation}

We may introduce the new phase angle $\beta$ by: 
\begin{equation}
\label{10}
\phi = \frac{\pi}{2} - \beta
\end{equation}

and then the amplitude (\ref{7}) takes the form: 
\begin{equation}
\label{11}
A^{2} = - \frac{1}{r + q \cos^{2}(\beta)} .
\end{equation}

The angle $\beta$ describes the tetragonal phase if it has value $\frac{\pi}{2}$. Near the 
phase transition from the tetragonal to the monoclinic phase the angle $ \beta < \frac{\pi}{2} $. 
From the equation (\ref{11}) we see that the amplitude A is almost independent on 
this angle. Near the phase transition from the tetragonal to the monoclinic phase 
we expect that the amplitude of the polarization vector does not change too 
much with temperature. This assumption corresponds with the experimental 
observations, \cite{19} to \cite{23}. Near the phase transition tetragonal to monoclinic phase 
the angle of the electric polarization changes mainly. We have found that at the 
transition temperature region the amplitude A is given by: 
\begin{equation}
\label{12}
A^{2} = - \frac{3 \alpha_{11} - \alpha_{12}}{2 \alpha_{111} - \alpha_{112} + O(\phi)^{2}} .
\end{equation}

As we can see from (\ref{12}) the amplitude A of the order parameter is weakly 
dependent on the angle $\beta$, or $\phi$. However it is nonzero above and below the 
transition temperature i.e. we may assume that it is almost independent on 
temperature near the phase transition from the tetragonal to the monoclinic 
phase. Then we may use the constant amplitude approximation $A = const. $ in 
our calculations near the phase transition temperature from tetragonal to the 
monoclinic phase. In this approximation we will assume that: 
\begin{equation}
\label{13}
A^{2} = - \frac{3 \alpha_{11} - \alpha_{12}}{2 \alpha_{111} - \alpha_{112} } 
\end{equation}

is constant. This approximation is corresponding to breaking the cubic symmetry 
for nonzero values of the angle $\phi$. Note that the free energy expansion 
(\ref{1}) for the cubic phase is a part of the free energy expansion describing the 
tetragonal to the monoclinic phase, however the free energy will contain contributions 
which correspond to the tetragonal phase symmetry now. The lowest 
order term (\ref{2}) has now the form: 
\begin{equation}
\label{14}
\delta F = \int dV \nu  A^{4} (\cos(\chi)^{2} - \sin(\chi)^{2})^{2} \sin(\phi)^{4}.
\end{equation}

Here the angle $\chi $ is the angle between the 1 and 2 components of the polarization. 
For very large $\nu$ constant this angle is 45 degrees in the monoclinic phase. We will assume that 
this angle has the value 45 degrees in the following. The discussion of a finite 
smaller value of the constant $\nu$ in (\ref{3}) will done elsewhere. 
From the equation (\ref{4}) and (\ref{5}) we obtain the following equation for the phase 
variable $\phi$: 
\begin{equation}
\label{15}
t^{4} [2 \alpha_{1} + 4 \alpha_{12} A^{2} + 2 (2\alpha_{112} + \alpha_{123}) A^{4}] +
\end{equation}
\[+ t^{2} [4 \alpha_{1} + 4 (\alpha_{11} + \alpha_{12} ) A^{2} + 8 \alpha_{112} A^{4}] + \]
\[ + [2 \alpha_{1} + 4 \alpha_{11} A^{2} + 6 \alpha_{111} A^{4}] = 0. \]

where $t = \tan(\beta)$. The monoclinic phase is realized when the free energy of it will be lower 
than that of the tetragonal phase. Let us discuss now solutions of the equation 
(\ref{15}). These solutions will correspond to extremes or inflection points of the free 
energy. The equilibrium state will be that solution which gives the lowest free 
energy. The equation is of the second order in the variable $t^{2} = \tan^{2}(\beta)$. The 
discriminant D of this equation is: 
\begin{equation}
\label{16}
D=[4\alpha_{1}+4(\alpha_{11}+\alpha_{12})A^{2}+8\alpha_{112}A^{4}]^{2}-
\end{equation}
\[-4[2\alpha_{1}+4\alpha_{12}A^{2}+2(2\alpha_{112}+\alpha_{123})A^{4}]
[2\alpha_{1}+4\alpha_{11}A^{2}+\]
\[+6\alpha_{111}A^{4}] \]

and should be real. Let us consider the real case. In the case of imaginary 
discriminant D the monoclinic phase does not exists. Let us again  consider the small amplitude A approximation. While this approximation is certainly correct 
near the cubic  tetragonal phase transition, it is not clear whether it is correct 
near the tetragonal to monoclinic phase transition. However we will assume that 
the difference $(2 \alpha_{11}-\alpha_{12})$ is small. The small constant amplitude approximation 
will be discussed numerically at the end of this section. In this approximation 
we may neglect the fourth order terms in A in the equation (\ref{16}). In this case the discriminant D has the form of the square of difference of the coefficients 
of the zeroth order (let us denote it as c) and of the fourth order (let us denote it as a) in the equation (\ref{15}). Then we have found that there exists a positive 
solution of the equation (\ref{15}) for $a > 0 > c$ and for $a < 0 < c$ for $t^{2}$ . In other 
cases of a and c there is a tetragonal phase realized. The solution of (\ref{15}) in the cases $a > 0 > c$ and $a < 0 < c$ has the form: 
\begin{equation}
\label{17}
\tan(\beta)^{2} = \mid \frac{c}{a} \mid 
\end{equation}

or: 
\begin{equation}
\label{18}
\cot(\phi)^{2} = \mid \frac{c}{a} \mid 
\end{equation}

which gives: 
\begin{equation}
\label{19}
\tan(\beta)^{2} = \mid \frac{\alpha_{1} + 2 \alpha_{12} A^{2}}{\alpha_{1} + 2 \alpha_{11} A^{2} } \mid 
\end{equation}

or: 
\begin{equation}
\label{20}
\cot(\phi)^{2} = \mid \frac{\alpha_{1} + 2 \alpha_{12} A^{2}}{\alpha_{1} + 2 \alpha_{11} A^{2} } \mid 
\end{equation}

Here the amplitude A is constant in temperature in this approximation, it is given by (\ref{13}). The equation (\ref{19}) may be rewritten for small angles $\beta$ in the form: 
\begin{equation}
\label{21}
\beta^{2} = \mid \frac{ T - T_{M}}{T - T^{N}} \mid 
\end{equation}

near the temperature $T_{M}$. 
The equation (\ref{20}) may be rewritten in the form: 
\begin{equation}
\label{22}
\phi^{2} = \mid \frac{ T - T^{N}}{T - T_{M} } \mid 
\end{equation}

near the temperature $T^{N}$ . 
Here the temperatures $T_{M}$  and $T^{N}$ are defined as: 
\begin{equation}
\label{23}
T_{M} = T_{0} - \frac{2 \alpha_{12} A^{2}}{\alpha_{1,0}}
\end{equation}

and: 
\begin{equation}
\label{24}
T^{N} = T_{0} - \frac{2 \alpha_{11} A^{2}}{\alpha_{1,0}}
\end{equation}

The difference of both temperatures is: 
\begin{equation}
\label{25}
T_{M} - T^{N} = - \frac{2 (\alpha_{12} - \alpha_{11} ) A^{2}}{\alpha_{1,0}}
\end{equation}

which is a positive or a negative quantity. For the phase transition from the 
tetragonal to monoclinic phase this difference is expected to be negative.

From the equation (\ref{22}) we see that the angle $\phi$ is zero at temperature $T^{N}$. 
This temperature is identified with the highest possible temperature  (the metastable) monoclinic phase may exist. Above 
this temperature there is the only solution of the Lagrange­Euler equations, the 
solution in which the only nonzero component of the polarization, $P_{3}$ , exists. 
This solution corresponds to the tetragonal phase. Below this temperature there 
exist nonzero components of the polarization, $P_{1}^{2} = P_{2}^{2} \ddagger P_{3}^{2} $ .

From the equation (\ref{22}) we see that the angle $\phi = \frac{\pi}{2}$ at temperature $T = T_{M}$ . This temperature is identified with the lowest possible temperature 
of the metastable monoclinic phase to exist. Above this temperature there may exist nonzero components of the polarization, $P_{1}^{2} = P_{2}^{2} \ddagger P_{3}^{2} $ .

From the equation (\ref{22}) we see that at the angle $\phi = \frac{\pi}{2}$ at temperature $T_{M} = \frac{T_{M} + T{N}}{2}$ 
the rhombohedral phase angle appears. Components $P_{1}^{2} = P_{2}^{2} = P_{3}^{2} $ 
are equal as concerning their amplitude at this temperature.

The temperature $T_{R}$ at which the rhombohedral phase angle value becomes present is given by: 
\begin{equation}
\label{26}
T_{R} - T^{N}=-\frac{(\alpha_{12}+\alpha_{11})A^{2}}{\alpha_{1,0}}.
\end{equation}

The calculations above were done for the approximation in which we neglected 
the terms of the order O(4) in the amplitude A. The angle $\phi$ development was calculated for temperature interval $T_{M} < T < T^{N} $ , i.e. for the interval from the highest possible temperature $T^{N}$ for which the monoclinic phase may exist, to 
the lowest temperature for which the monoclinic phase may exists. The temperature 
at which the rhombohedral angle value of $\phi$ realizes is at the middle 
of this interval.

We have calculated the numerical values of the amplitude $A^{2}$ for the concentrations 
50/50, 40/60, 30/70 and 20/70 using the numerical values of the 
constants in the Landau expansion for the free energy from \cite{19} to \cite{23} . We 
have found that it has a negative value ­3.690 in the corresponding units for the 
50/50 ratio, and the positive value +0.834 for the 40/60 ratio increasing further 
its value to +6.039 and +11.880 for 30/70 and 20/80 ratios respectively. For the 
10/90 ratio it has a negative value ­4.507. We have found that the constant small 
amplitude approximation near the boundary 50/50 is a good approximation for 
PZT and that $A^{2}$ is a small positive parameter in this region where it is changing its sign. The 
amplitude A has nonzero value at tetragonal phase. The phase transition from 
tetragonal to monoclinic phase is of the first order near to the second order. 
Thus the negative numerical value of the amplitude $A^{2}$ for the monoclinic phase for the concentrations 
50/50 means that the monoclinic phase does not realize for this concentration. 
Note that the amplitude $A^{2}$ is calculated for the monoclinic phase. Decreasing 
temperature the free energy of the monoclinic phase is expected to increase, 
and at some point it may become larger than that of the rhombohedral phase.

\section{Transition Temperature Region between the 
Monoclinic and the Tetragonal Phase.}

Let us consider the free energy of the tetragonal and the monoclinic phase. 
Comparing them for the equilibrium states we obtain the transition temperature 
between the monoclinic and the tetragonal phase. The later phase is that phase 
for which $P_{1}^{2} = P_{2}^{2} = 0 $ and $P_{3}^{2} \ddagger 0 $. The former phase is that phase for which 
$P_{1}^{2} = P_{2}^{2} \ddagger 0 $ and $P_{3}^{2} = 0 $, 
the components 1 and 2 of the polarization vector 
are different from the component 3. The transition temperature between the 
tetragonal phase and the cubic phase, in which the polarization is zero, is: 
\begin{equation}
\label{27}
T_{T-C} = T_{0} + \frac{ \alpha_{11}^{2}}{4 \alpha_{1,0}\alpha_{111}}
\end{equation}

As we see this temperature is different from temperature $T_{0}$ which contains the coefficient $\alpha_{1}$ . The phase transition is of the second order near the composition 
50/50 and of the first order near the second order for higher concentrations, see \cite{19} to \cite{23}.
One can use the concentration dependence of the coefficients $\alpha$ in the free energy 
expansion and find how this transition temperature, and also other quantities, 
is changing with the concentration, \cite{19} to \cite{23}. This is not our aim here. 
The free energy of the monoclinic phase will be compared with that of the 
tetragonal phase. The free energy of the monoclinic phase is given within our 
approximation of $\nu$ large by: 
\begin{equation}
\label{28}
F = \int dV [ \alpha_{1} A^{2} \frac{1 + 2 t_{0}^{2}}{1 +  t_{0}^{2}}+
\end{equation}
\[ + \alpha_{11} A^{4} \frac{1 + 2 t_{0}^{4}}{(1 +  t_{0}^{2})^{2}} + 
\alpha_{12} A^{4} \frac{(2 +  t_{0}^{2})t_{0}^{2}}{(1 +  t_{0}^{2})^{2}}  + \]
\[ + \alpha_{111} A^{6} \frac{1 + 2 t_{0}^{6}}{(1 +  t_{0}^{2})^{3}} + 
\alpha_{112} A^{6} \frac{t_{0}^{2} + t_{0}^{4} + t_{0}^{6} }{(1 +  t_{0}^{2})^{3}}  + \alpha_{123} A^{6} \frac{ t_{0}^{6} }{(1 +  t_{0}^{2})^{3}}]   \]

Here $t_{0} = \tan(\phi)$. The free energy of the tetragonal phase has within our approximation the form: 
\begin{equation}
\label{29}
F = \int dV [ \alpha_{1} A^{2} + \alpha_{11} A^{4} + \alpha_{111}A^{6}].
\end{equation}

In our approximation we have in the tetragonal phase $P^{2}_{3} = A^{2}$ . The amplitude $A^{T}$ in the tetragonal phase is given for PZT in the range near MPB by: 
\begin{equation}
\label{30}
A^{T 2} = \frac{-2 \alpha_{11} + \sqrt{(2 \alpha_{11})^{2} - 12 \alpha_{111} \alpha_{1}}}{6 \alpha_{111}}.
\end{equation}

The transition temperature from the tetragonal to the monoclinic phase is 
found from free energies of the tetragonal phase and of the monoclinic phase comparing them.
At this temperature they are equal. Above this temperature the free energy 
of the tetragonal phase is lower than that of the monoclinic phase, below this 
temperature the free energy of the tetragonal phase is higher than that of the 
monoclinic phase. For the monoclinic phase we will use the small constant 
amplitude approximation as described above at this region of temperatures. 
Then we obtain for the transition temperature the equation: 
\begin{equation}
\label{31}
3D - \frac{D \sqrt{D}}{\alpha_{11}} = 4 \alpha_{11}^{2} - 2 \alpha_{12} A^{T 4} \frac{108 \alpha_{111}^{2}}{\alpha_{11}}
\end{equation}

where the discriminant D is given by: 
\begin{equation}
\label{32}
D = (2 \alpha_{11})^{2} - 12 \alpha_{111}\alpha_{1}
\end{equation}

Note that in between 40/60 and 50/50 concentrations there is present a 
large change of the constant $ y \equiv \frac{6 \alpha_{111} \mid \alpha_{1}\mid }{4\alpha_{11})^{2}}$
where in $\alpha_{1}$ we take the temperature 
for the transition from the tetragonal to monoclinic phase as 358K. For the 
concentration 40/60 this constant has its value 1.179 and for the concentration 
50/50 the value 60.387 . The discriminant D above can be written as $
(2 \alpha_{11})^{2}(1 + 2y)$. Nearby the MPB the constant y is expected to be large. Then we obtain 
from (32) that the transition temperature $T_{T-M}$ from the tetragonal to the monoclinic phase is given by: 
\begin{equation}
\label{33}
T_{T-M} = T_{0} - \frac{ \alpha_{12}^{2}}{3 \alpha_{1,0} \alpha_{111}}(1 + \alpha_{12} A^{T, 4} \frac{108 \alpha_{111}^{2}}{4 \alpha_{11}^{3}})
\end{equation}

To estimate the numerical value for this transition temperature we need 
the values of the constants in (\ref{33}) for the concentrations near MPB for concentrations 
near 48/52. These are not available. For 50/50 concentrations we obtained from (\ref{33}) that 
temperature $T_{T-M}$ in (\ref{33}) is negative. It means that at this concentration there 
is no monoclinic phase. This corresponds with experimental findings up to now. 
For concentrations 40/60 our approximation of a small constant amplitude is 
not valid. Moreover the rhombohedric phase is preferred at this concentration 
as that phase which evolves from the cubic phase decreasing temperature.

\section{Dielectric Response Near the Phase Transition.}

The dielectric response of PZT in the tetragonal and monoclinic phases will 
be calculated from the free energy given by (\ref{1}) and (\ref{2}) in the approximation of large $\nu$ applying an external 
electric field E in the direction 3. The external field will interact with 
the polarization, and this interaction gives a contribution 
$-{\bf E.P }$ to the free 
energy density. One can ask how PZT responses to the field which is  parallel and perpendicular to the tetragonal phase polarization direction in the monoclinic phase. One can also ask how PZT responses to the electric 
field in the monoclinic phase where there exist polarization components (1,2) 
which are absent in the tetragonal phase. 
Dielectric response will be calculated from the total free energy for the tetragonal phase.

Let us first consider how PZT responses to the fields which are parallel to the tetragonal phase polarization axis in the tetragonal phase and in the 
monoclinic phase. Let us assume that there is a nonzero electric field $E_{3}$ in 
the direction (001) in pseudo cubic notation.

In the tetragonal phase the field 
parallel to the polarization induces changes in the amplitude of the polarization 
vector, the monoclinic angle $\phi$ is zero and remains zero for this field. The change 
$\delta A$ in the amplitude A induced by this field is: 
\begin{equation}
\label{34}
\delta A = - \frac{1}{4A^{2}\sqrt{D}} E_{3}.
\end{equation}

From the equation (ref{34}) we see that there is present a diverging behaviour from the cubic to tetragonal phase in the susceptibility $\epsilon_{33}$. Note that D is defined in the equation (\ref{32}) 
and the amplitude $A$ in the tetragonal phase is given by (\ref{30}). This field induces a change of the polarization in the direction (001) and does not induce a change of the polarization in the direction (100) or (010).

Near the transition from tetragonal to monoclinic phase the amplitude A in the monoclinic phase is 
more­less constant, we will use this approximation.

The angle $\phi$ is nonzero in 
the monoclinic phase, thus we may expect that nonzero electric field $E_{3}$ induces 
a change not only in the direction (001) but also in the directions (100) and 
(010). From the equations (\ref{4}) and (\ref{5}) we find that the polarization components 
are related by the equation: 
\begin{equation}
\label{35}
1 + r(P_{1}^{2} + P_{3}^{2}) + q P_{1}^{2}=
\frac{2 E_{3}}{(P_{3}^{2} - P_{1}^{2})P_{3}}.
\end{equation}

In the polar order parameter space $ A, \phi $ we obtain from (\ref{35}): 
\begin{equation}
\label{36}
1 + r A^{2} + q A^{2} \sin^{2}(\phi) = 
\frac{2 E_{3}}{A^{3}\cos(\phi)\cos^{2}(2 \phi)}.
\end{equation}

As we can see from the equation (\ref{36}) the amplitude change induced by the 
external field $E_{3}$ is coupled to the angle $\phi$ change induced by this field.

We have found that the dielectric susceptibility $\epsilon_{33}$ in the monoclinic phase is given in the monoclinic phase within our approximations by a contribution due to the induced change of the angle: 
\begin{equation}
\label{37}
\epsilon^{ph}_{33} = \frac{2(1 + t^{2}_{0})^{2})}{q \sqrt{\frac[9]{-1}{r}}(2 - 7 t^{2}_{0} + t^{4}_{0})}
\end{equation}

and by a contribution due to the induced change of the amplitude: 
\begin{equation}
\label{38}
\epsilon^{am}_{33} = \frac{-2r}{ 3 \sqrt{1 + t^{2}_{0}}}.
\end{equation}

The total dielectric susceptibility $\epsilon_{33}^{M}$ for the monoclinic phase is given within the constant amplitude 
approximation by the sum of (\ref{37}) and (\ref{38}).

We have found that the dielectric susceptibility $\epsilon_{13}$ in the monoclinic phase is given within 
the constant amplitude approximation in the monoclinic phase by: 
\begin{equation}
\label{39}
\epsilon_{13} = \epsilon_{33} [ \frac{\mid t_{0}\mid }{ \sqrt{1 + t^{2}_{0}}} +
\end{equation}
\[ + \frac{3(1 + t^{2}_{0})^{2}}{q \sqrt[9]{\frac{-1}{r}}\mid t_{0}\mid (2 - 7 t^{2}_{0} + t^{4}_{0})}]. \]

Here we denoted $t_{0}=\tan(\phi)$ and $\epsilon_{33}^{T}$:
\begin{equation}
\label{40}
\epsilon_{33}^{T} = - \frac{2r}{3}.
\end{equation}

It is easy to find the susceptibility of the tetragonal phase as we described it above, see also \ref{19} to \ref{23}. It is temperature dependent and diverging at the trantision temperature from the cubic to the tetragonal phase for the second order phase transition, and almost diverging for the first order phase transition near the second order.

Near the tetragonal to monoclinic phase transition temperature we have 
found that dielectric susceptibility $\epsilon_{13}$ is zero in the tetragonal phase: 
\begin{equation}
\label{41}
\epsilon_{13} = 0.
\end{equation}

Near tetragonal to monoclinic phase transition temperature we have found in the monoclinic phase the dielectric susceptibility $\epsilon_{13}$ to be given by: 
\begin{equation}
\label{42}
\epsilon_{13} = \epsilon_{33}^{T} [\frac{3}{q \sqrt[9]{\frac{-1}{r}} \mid t_{0} \mid}]
\end{equation}

The dielectric susceptibility $ \epsilon_{13}$ diverges at the temperature $ T^{N} $ .

As we can see the susceptibility $\epsilon_{33}$ in the direction of the applied electric 
field is nonzero in the tetragonal phase and in the monoclinic phase. Decreasing 
temperature it is decreasing in the monoclinic phase in a slower way than 
in the tetragonal phase. The slope of decreasing susceptibility in tetragonal 
phase changes at this transition temperature to smaller one. This behavior is seen in the 
experiment, see \cite{12}. 
The susceptibility $\epsilon_{33}$ has its value at the temperature $T_{N}$ : 
\begin{equation}
\label{44}
\epsilon_{33}=\epsilon_{33}^{T}[\frac{1}{\sqrt{2}}+ \frac{3}{q\sqrt[9]{\frac{-1}{r}}}]
\end{equation}

which leads to the value $\frac{\epsilon_{33}}{\epsilon_{33}^{T}}$
= 0.906 at this temperature. 
As we can also see the susceptibility $\epsilon_{13}$ is zero in the tetragonal phase 
and in the monoclinic phase it is either increasing either decreasing function 
in temperature depending on the sign in the equation (\ref{39}). For the 40/60 
concentration ratio we have found that it is a negative quantity. Applying the 
field $E_{3}$ in the monoclinic phase there is a contribution to the susceptibility 
due to the change of the angle $\phi$ in such a way that the polarization vector is 
tilted to the tetragonal axis (001), thus the $\epsilon_{13}$ susceptibility is expected to be negative. Note that the constant r is independent on 
temperature in our approximation, however it is dependent on the PZT solid 
solution concentrations ratio. So there is a jump in this component of the 
susceptibility. There is also a jump in the susceptibility component $\epsilon_{13}$ present for the crystal case studied.

\section{PZT Ceramics ­ Their Microcomposite Properties and Dielectric Response.}

The ceramic PZT is composited from crystallites which have polar axis oriented 
in different directions and which have different shape, and from a free 
volume. There exists a possibility of coexistence of crystallites with different 
phases tetragonal and rhombohedral at and nearby MPB which leads us to the 
question whether such coexistence does not lead to a possible enhancement or decrease of the 
dielectric response \cite{27} in this material. The effect of different phases in different 
crystallites in PZT was considered in \cite{13}. According to this work we can 
consider PZT as a microcomposite. Let us calculate the effective response of 
the PZT microcomposite. It consists of grains which have the tetragonal axis 
(001) oriented in a different ways. The effective dielectric susceptibility of the 
microcomposite $\epsilon_{eff}$ is found within the Effective Medium Approximation \cite{27} from: 
\begin{equation}
\label{45}
x\frac{\epsilon_{eff}- \epsilon_{1}}{2 \epsilon_{eff} + \epsilon_{1}} + (1 - x) \frac{\epsilon_{eff} - \epsilon_{2}}{2 \epsilon_{eff} + \epsilon_{2}} = 0.
\end{equation}

Here the susceptibilities $\epsilon_{1}$ and $\epsilon_{2}$ are susceptibilities for two types of materials (grains) in the microcomposite. The concentration x is a concentration 
of the first type of the material, the concentration 1­x is a concentration of the 
second type of the material. Let us assume that the first type of material is 
that part of grains which has its tetragonal axis oriented in the direction of the 
field. Let us further assume that the second type of material is that part of 
grains which has its tetragonal axis oriented perpendicular to the direction of 
the field. This is a simplified model of the PZT ceramics, we expect that it 
gives a qualitative picture of the microcomposite response. The effective $\epsilon_{eff}$  
constant is given in the tetragonal phase, where the off diagonal dielectric susceptibility 
is zero, given by the $\epsilon_{33}$ constant. In the monoclinic phase the 
offdiagonal dielectric constant is large, diverging at $T^{N}$ . It can be easily found 
from (\ref{45}) that the effective dielectric constant for concentrations x of the first 
type of material above the critical concentration for percolation of spheroidal 
grains $x_{c} = \frac{2}{3}$ is given by: 
\begin{equation}
\label{46}
\epsilon_{eff} = \frac{\epsilon_{1}}{3(x  -  x_{c})}.
\end{equation}

Thus it is a finite response, determined by the first type of grains. Note that 
the closer we are to the critical concentration the larger is this effective dielectric 
constant. It is this behavior which corresponds to the observed behavior, see in 
\cite{12}. Below the critical concentration the effective: 
\begin{equation}
\label{47}
\epsilon_{eff} = \frac{3 \epsilon_{2}}{2}(x  -  x_{c}).
\end{equation}

This effective dielectric constant is large, almost diverging depending on 
divergence of the dielectric constant $\epsilon_{13}$.

In the tetragonal phase we have found that the effective susceptibility is 
given by: 
\begin{equation}
\label{48}
\epsilon_{eff} = \epsilon_{33}.
\end{equation}

\section{Mechanical Response Near the Phase Transition.}

Let us consider a crystal of PZT which is under the uniaxial stress $\sigma $. We will use symmetry considerations of Haun \cite{19} to \cite{23} to find a Gibbs free energy expansion. However in this case we need to consider the coupling between the elastic tensor components (strain) and the polarization vector components. The nonzero 
uniaxial stress $ \sigma_{3}$ interacts with the $ u_{3} $ component of the elastic tensor. We are using notation 1 to 6 for tensor components of the strain tensor u. This 
component interacts with other components of the elastic tensor (strain) and with the 
polarization components in a way described by the part of the Gibbs free 
energy which has the same form as the part used by Haun et al. \cite{19} to \cite{23} for the Gibbs free energy with stress $ \sigma_{ij} $, i,j=1, 2, 3. The thermodynamic potential $ \delta F  $ (the Gibbs free energy) expansion in strain is used in the same way as in Landau and Lifshitz book \cite{23LL}: 
\begin{equation}
\label{49}
\delta F = \int dV [ \frac{1}{2} C_{11}(u_{1}^{2} + u_{2}^{2} + u_{3}^{2}) + C_{12}(u_{1}u_{2} + u_{2}u_{3} + u_{3}u_{1}) +
\end{equation}
\[ + \frac{1}{4}C_{44}(u_{4}^{2} + u_{5}^{2} + u_{6}^{2}) + q_{11}(u_{1}P_{1}^{2} + u_{2}P_{2}^{2} + u_{3}P_{3}^{2}) + \]
\[ + q_{12}(u_{1}(P_{2}^{2} + P_{3}^{2}) + u_{2}(P_{3}^{2} + P_{1}^{2}) + u_{3}(P_{1}^{2} + P_{2}^{2})) + \]
\[ + q_{44}(u_{4}P_{2}P_{3} + u_{5}P_{3}P_{1} + u_{6}P_{1}P_{2}) - u_{3} \sigma_{3}]. \]

The uniaxial stress in the direction 3, $\sigma_{3}\equiv \sigma$ 
in the following. The elastic 
tensor components (strain) have their values from the following equations: 
\begin{equation}
\label{50}
C_{11}u_{3} + C_{12}(u_{1}+u_{2}) + q_{11}P_{3}^{2} + 
q_{12} (P_{1}^{2} + P_{2}^{2}) - \sigma = 0,
\end{equation}

and: 
\begin{equation}
\label{51}
C_{11}u_{1} + C_{12}(u_{2}+u_{3}) + q_{11}P_{1}^{2} + 
q_{12} (P_{2}^{2} + P_{3}^{2}) = 0,
\end{equation}

and: 
\begin{equation}
\label{52}
C_{11}u_{2} + C_{12}(u_{3}+u_{1}) + q_{11}P_{2}^{2} + 
q_{12} (P_{3}^{2} + P_{1}^{2}) = 0.
\end{equation}

When the coupling between the polarization and the strain is absent, i.e. 
when $q_{12} = q_{11} = 0$, then the deformations in the 1 and 2 directions have the value, 
$u_{1} = u_{2} = - u_{3} \frac{2C_{12}}{C_{11} + C_{12}} $, and in the direction 3 it has the value $u_{3} =  \frac{\sigma}{C_{11} - \frac{2 C_{12}^{2}}{C_{11} + C_{12}}} $. When the coupling is present, $q_{12} = \ddagger  0 $ and $q_{11} \ddagger  0$, then we obtain for the component $u_{3}$: 
\begin{equation}
\label{53}
u_{3} = \frac{1}{C_{11}(1 - \frac{2 C_{12}^{2}}{C_{11}(C_{11} + C_{12})})}[\sigma - q_{11}P_{3}^{2} - q_{12} (P^{2}_{1} + P^{2}_{2})+
\end{equation}
\[ + \frac{C_{12}}{C_{11} + C_{12}} (q_{11}(P_{1}^{2} + P_{2}^{2}) + q_{12}(P_{1}^{2} + P_{2}^{2} + 2 P_{3}^{2})) ] . \]

As we see from (\ref{53}) the strain tensor component $u_{3}$ depends on the stress $\sigma $ and on the components of the polarization vector. In the cubic phase $P_{1} = 
P_{2} = P_{3} = 0$ and the tensor component $u_{3}$ is given by: 
\begin{equation}
\label{54}
u_{3} = \frac{1}{C_{11}(1 - \frac{2 C_{12}^{2}}{C_{11}(C_{11} + C_{12})})}\sigma. 
\end{equation}

In the tetragonal phase $P_{1} = P_{2} = 0$ and $P^{2}_{3} > 0$ the elastic tensor component $u_{3}$ is given by: 
\begin{equation}
\label{55}
u_{3} = \frac{1}{C_{11}(1 - \frac{2 C_{12}^{2}}{C_{11}(C_{11} + C_{12})})}[ \sigma - (q_{11} - \frac{2C_{12}}{C_{11} + C_{12}})P_{3}^{2}].
\end{equation}

In the monoclinic phase $P^{2}_{1} = P^{2}_{2} \ddagger P^{2}_{3} = 0 $ the elastic tensor component $u_{3}$ has the form in the $A, \phi $ space: 
\begin{equation}
\label{56}
u_{3} = \frac{1}{C_{11}(1 - \frac{2 C_{12}^{2}}{C_{11}(C_{11} + C_{12})})}[ \sigma - A^{2}(q_{11} \frac{1}{1 + t_{0}^{2}} + 2q_{12}
\frac{t_{0}^{2}}{1 + t_{0}^{2}}) +
\end{equation}
\[ + \frac{2C_{12}}{C_{12} + C_{11}}(q_{11}A^{2}\frac{t_{0}^{2}}{1 + t_{0}^{2}} + q_{12}A^{2} )]. \]

In the tetragonal and in the monoclinic phase we have the change of the 
strain tensor $\delta u_{3}$ induced by the (small) uniaxial stress $\sigma$, for zero coupling between     the strain and the polarization: 
\begin{equation}
\label{57}
\delta u_{3} = \frac{1}{C_{11}(1 - \frac{2 C_{12}^{2}}{C_{11}(C_{11} + C_{12})})} \sigma 
\end{equation}

In the tetragonal and in the monoclinic phase we have the change of the 
strain tensor components $\delta u_{1}$ and $\delta u_{2}$ induced by the uniaxial stress $\sigma$ for zero  coupling between the strain and the polarization: 
\begin{equation}
\label{58}
\delta u_{1} = \delta u_{2} = - \frac{\sigma}{C_{11} + \frac{2 C_{12}^{2}}{C_{11} + C_{12}}}  \frac{C_{12}}{C_{11} + C_{12}}. 
\end{equation}

In the previous calculations of the strain tensor we have neglected the influence 
of the coupling between the strain and the polarization and we obtained the relations (\ref{57}) and (\ref{58}). We see from them that the strain tensor components are temperature independent.
However this coupling has to be taken into account, 
in experiments \cite{12} we see that the strain is temperature dependent. From the equations (\ref{57}) and (\ref{58})
we define $x_{33}$, $x_{13}$, and $x_{23}$ by $u_{3} = x_{33} \sigma $, $u_{1} = x_{13} \sigma$ and $u_{2} = x_{23} \sigma$. 
Here $x_{13} = x_{23}$ .
Taking into account the coupling between the strain and the polarization,  i.e. the nonzero coefficients 
$q_{11}$, $q_{12}$ and $q_{44}$, the  $\alpha_{1}$ coefficient renormalizes to $\alpha_{1,1}$ and $\alpha_{1,3}$ coefficients, where: 
\begin{equation}
\label{59}
\alpha_{1,1} = \alpha_{1} + (q_{11} x_{13} + q_{12} (x_{13} + x_{33} ))
\end{equation}

and where: 
\begin{equation}
\label{60}
\alpha_{1,3} = \alpha_{1} + (q_{11} x_{33} + q_{12} (x_{13} + x_{23} ))
\end{equation}

In the monoclinic phase, taking into account the coupling between the electric 
polarization vector and the elastic tensor (strain) components, we obtain the 
following Lagrange Euler equations for the components $P_{1}$ and $P_{3}$ : 
\begin{equation}
\label{61}
4 \alpha_{1,1} + 4 (2 \alpha_{11} + \alpha_{12})P_{1}^{2} + 4 \alpha_{12}P_{3}^{2} + 12 (\alpha_{111} + \alpha_{112})P_{1}^{4} +
\end{equation}
\[ + 4 \alpha_{112} P_{3}^{4} + 4 (\alpha_{112} + \alpha_{123})P_{1}^{2}P_{3}^{2} = 0 \]

and: 
\begin{equation}
\label{62}
2 \alpha_{1,3} + 4 \alpha_{11}P_{3}^{2} + 4 \alpha_{12}P_{1}^{2} + 
6 \alpha_{111} P_{3}^{4} +
\end{equation}
\[ 8 \alpha_{112})P_{1}^{2}P_{3}^{2} +  2(2 \alpha_{112} + \alpha_{123} )P_{1}^{4}= 0 \]

From the equations (\ref{61}) and (\ref{62}) we obtain a relation between the stress 
and the polarization vector, neglecting terms $A^{4}$ in the small amplitude approximation, expressed in the polar coordinates $A, \phi$: 
\begin{equation}
\label{63}
Q \sigma - 4 (2 \alpha_{11} + \alpha_{12})A^{2} \cos(\phi)= 0,
\end{equation}

where the coefficient Q is defined by: 
\begin{equation}
\label{64}
Q = 4(q_{11}x_{13} + q_{12}(x_{33} + x_{23})) - 4(q_{11}x_{33} + Q_{12}(x_{23} + x_{13})). 
\end{equation}

As wee see from the equations (\ref{63}) and (\ref{64}) the stress is related to the amplitude 
A and to the angle $\phi$ when the coefficients $q_{11}$ and $q_{12}$ are nonzero. 
The stress induces a change in the angle $\phi$ which leads to a temperature independent renormalization of the $\frac{\delta u^{M}_{3}}{\delta \sigma}$ component of the inverse compliance in 
our approximation. For the constant amplitude approximation in (\ref{64}) we obtain for the $\frac{\delta u^{M}_{3}}{\delta \sigma}$ component of the inverse compliance in this case within our approximation: 
\begin{equation}
\label{65}
\frac{\delta u^{M}_{3}}{\delta \sigma} = \frac{1}{C_{11}} \frac{1}{1 - 2 \frac{C_{12}^{2}}{C_{11}(C_{11} + C_{12})}} [1 - 
(2 q_{12} - q_{11})\frac{Q}{8(\alpha_{12} - 2 \alpha_{11})} + .
\end{equation}
\[ + \frac{2 C_{12}}{C_{11} + C_{12}} q_{11})\frac{Q}{8(\alpha_{12} - 2 \alpha_{11})}]. \]

From the equations (\ref{63}) and (\ref{64}) it follows however that the stress induces 
a change in the amplitude A such that it leads to a temperature dependent 
renormalization of the $\frac{\delta u^{M}_{3}}{\delta \sigma}$ component of the inverse compliance in our approximation. 
For the constant angle $\phi$ in (\ref{64}) we obtain for the $\frac{\delta u^{M}_{3}}{\delta \sigma}$ component of 
the inverse compliance in our approximation: 
\begin{equation}
\label{66}
\frac{\delta u^{M}_{3}}{\delta \sigma} = 
\frac{1}{C_{11}} \frac{1}{1 - 2 \frac{C_{12}^{2}}{C_{11}(C_{11} + C_{12})}}
[1 - Q \frac{(q_{11} + q_{12}t_{0}^{2} + \frac{2C_{12}}{C_{11} + C_{12}}(q_{11})t_{0}^{2} + q_{12} ) }{4 (2 \alpha_{11} -  \alpha_{12})(1 - t_{0}^{2})}].
\end{equation}

Thus a temperature dependent strain component 
in the monoclinic phase is found due to a change in the amplitude of the polarization vector and due to the coupling between the strain and the polarization.

We can write the quantity $ \frac{\delta u^{M}_{3}}{\delta \sigma} $ in a general form appropriate for fitting experimental curves: 
\begin{equation}
\label{67}
\frac{\delta u^{M}_{3}}{\delta \sigma} = 
\frac{1}{C_{11}} \frac{1}{1 - 2 \frac{C_{12}^{2}}{C_{11}(C_{11} + C_{12})}}
[1 + \tau_{0} + \frac{\tau_{1} + \tau_{2}t_{0}^{2}}{1 -t_{0}^{2}}].
\end{equation}

We see from the equation (\ref{66}) that the quantity $ \frac{\delta u^{M}_{3}}{\delta \sigma} $ 
is finite as the temperature T increases to the temperature $T^{N}$, the temperature 
at which the monoclinic phase becomes unstable for higher temperatures.
With decreasing temperature the inverse of the $\frac{\delta u^{M}_{3}}{\delta \sigma}$ function, which is measured in experiments \cite{11} and \cite{12}, becomes 
larger in the monoclinic phase as can be seen from (\ref{66}).

In the tetragonal phase the inverse of the $\frac{\delta u^{M}_{3}}{\delta \sigma}$,which is connected to the Young modulus, is calculated from its 
free energy expansion taking into account the strain contribution, the strain 
 polarization component coupling and the stress strain contribution. We 
obtained that the strain component $u_{3}$ has the form for nonzero uniaxial stress: 
\begin{equation}
\label{72}
u_{3} = \frac{1}{C_{11} - \frac{2C_{12}^{2}}{C_{11}(C_{11} + C_{12})}} [\sigma - 
\end{equation}
\[ - (q_{11} - q_{12} \frac{ C_{12}}{C_{11} + C_{12}} \frac{1}{6 \alpha_{111}}(-4 \alpha_{11}^{,} + \sqrt{(4 \alpha_{11}^{,})^{2} - 12 \alpha_{111}(\alpha_{1} + \sigma \rho_{1})}))]. \]

Here $\alpha_{11}^{,} = \alpha_{11} - q_{11} (q_{11} - q_{12} \frac{2 C_{12}}{C_{11} + C_{12}}) + q_{12}(\frac{C_{12}}{C_{11} + C_{12}} \frac{2}{C_{11} - \frac{2C_{12}^{2}}{C_{11} -C_{12}}})(q_{11} - q_{12} \frac{2C_{12}}{C_{11} + C_{12}})$

and $ \rho_{1} = \frac{1}{C_{11} - \frac{2C_{12}^{2}}{C_{11} + C_{12}}}(q_{11} - q_{12}\frac{2C_{12}^{2}}{C_{11} + C_{12}})+ q_{11} + q_{12}(- \frac{C_{12}}{C_{11} + C_{12}}\frac{2}{C_{11} - \frac{2C_{12}^{2}}{C_{11}+C_{12}}})$.

With increasing temperature T the 
inverse of the $\frac{\delta u^{M}_{3}}{\delta \sigma}$, see (\ref{72}), which is measured in experiments, increases in tetragonal phase.

Temperature dependence of the 
inverse of the $\frac{\delta u^{M}_{3}}{\delta \sigma}$  below the transition temperature from monoclinic to tetragonal phase is such that it increases as temperature T decreases. 
The inverse of the $\frac{\delta u^{M}_{3}}{\delta \sigma}$ quantity is increasing with increasing temperature above the transition temperature from monoclinic to tetragonal phase. This behaviour is expected to occur only in the neighbourhood of the Morphotropic Phase Boundary, where we expect that $q_{11} + 2 q_{12}$ is positive. Here $q_{11} \approx Q_{11} $ and $q_{12} \approx Q_{12} $, the quantity $Q_{11}$ and the quantity $Q_{12}$ are dependent on the composition of PZT, see in \cite{21}.
This behaviour of the mechanical response function $\frac{\delta u^{M}_{3}}{\delta \sigma}$ corresponds qualitatively to the known experimentaly measured dependences, see in \cite{11} and \cite{12}.

\section{Summary.}

A phase transition between the tetragonal phase and the monoclinic phase was 
described within the Landau theory of phase transitions of the second order and 
of the first order near the second order. We have found that the transition from 
the tetragonal to monoclinic phase is driven by the electric polarization vector. 
So it is not necessary to introduce an artificial order parameter for this phase 
transition. The transition is of the first order near the second order. The tetragonal 
phase and the monoclinic phase equilibrium states were found from the free energy 
expansion for the tetragonal phase symmetry. We have used the constant amplitude approximation to 
describe the monoclinic phase and the phase transition from 
tetragonal to monoclinic phase. Such an approximation corresponds to the fact 
that the polarization vector does not change its amplitude too much during this 
phase transition, which was observed in \cite{12}. The transition is studied in the small 
constant amplitude approximation because a numerical estimate of the amplitude 
using the numerical values data for PZT free energy expansion coefficients 
in the cubic phase, as found by Haun et al. \cite{19} to \cite{23}, has shown that it is a 
small quantity in dimensionless units. To estimate the numerical value for the transition temperature we used
the values of the constants in (\ref{33}) for the concentrations near MPB for concentrations 
near 48/52. For 50/50 concentrations we obtained that 
temperature $T_{T-M}$ in (\ref{33}) is negative, at this concentration there is no monoclinic phase. This corresponds with experimental findings up to now.

Applying an external electric field in the 
tetragonal direction and an external uniaxial stress in this direction we studied in 
this paper dielectric and mechanical responses of the PZT crystal in the tetragonal 
and in the monoclinic phases. To explain observed dependencies in \cite{11} and 
\cite{12} of the static effective dielectric susceptibility we had to take into account 
that samples on which dielectric response was measured were either polycrystals 
or microcomposites. A model for such a microcomposite was formulated in \cite{26}. 
We have found that large off diagonal dielectric response present in the monoclinic 
phase near the transition from the tetragonal phase due to its divergence 
at the temperature of the stability of the monoclinic phase is not seen 
in the experiments because of the polycrystalline and of microcomposite nature 
of samples in which the concentration of the components with off diagonal 
dielectric susceptibility component are below the percolation threshold. Note 
that according to our theory one should observe at least a  kink in this component 
in a pure crystal. We have succeeded to explain qualitatively the observed 
temperature behavior of the effective dielectric response, the formulas describing 
quantitative this behavior have been found in this paper, however a fit of 
them to experimentally observed curves will be done elsewhere \cite{11}. 
We have 
studied the effective dielectric response of PZT polycrystals and ceramics using 
the Effective Medium Approximation.

The mechanical response was also studied. The mechanical response tensor 
33 and 13 (in the shortened notation 1­6) was found for a crystal for the tetragonal phase and the monoclinic phase. Its temperature behaviour corresponds to that observed in the experiment, there is a well around the transition temperature.

\section*{Acknowledgment}

One of the authors (O.H.) wishes to express his sincere thanks to the Istituto 
dei Sistemi Complessi, CNR, Roma for kind hospitality and financial support of his stay.

\end{document}